\documentclass[fleqn,usenatbib]{mnras}

\usepackage[T1]{fontenc}
\usepackage{ae,aecompl}

\usepackage{graphicx}	
\usepackage{amsmath}	
\usepackage{amssymb}	

\title[Heating of Stellar Coronae]
{Subresolution Activity in Solar and Stellar Coronae from Magnetic Field Line Tangling}

\author[A. F. Rappazzo et al.]{
A. F. Rappazzo,$^{1}$\thanks{E-mail: rappazzo@ucla.edu (AFR)}
R. B. Dahlburg,$^{2}$
G. Einaudi$^{3}$
and M. Velli,$^{1}$
\\
$^{1}$Department of Earth, Planetary and Space Sciences,
UCLA, Los Angeles, CA 90095, USA\\
$^{2}$Laboratory for Computational Physics and Fluid Dynamics, Naval Research Laboratory, Washington, DC 20375, USA\\
$^{3}$Department of Physics and Astronomy, George Mason University, Fairfax, VA 22030, USA
}

\date{Accepted 2018 April 30. Received 2018 April 16; in original form 2018 February 22}

\pubyear{2018}

\usepackage{etoolbox}
\makeatletter
\patchcmd\@combinedblfloats{\box\@outputbox}{\unvbox\@outputbox}{}{%
	\errmessage{\noexpand\@combinedblfloats could not be patched}%
}%
\makeatother
\begin{document}
\label{firstpage}
\pagerange{\pageref{firstpage}--\pageref{lastpage}}
\maketitle
\begin{abstract}
The heating of coronal loops is investigated to understand the observational consequences in terms of the thermodynamics and radiative losses from the Sun as well as the magnetized coronae of stars with an outer convective envelope. The dynamics of the Parker coronal heating model are studied for different ratios of the photospheric forcing velocity timescale $t_p$ to the Alfv\'en crossing time along a loop $t_A$. It is shown that for $t_p/t_A \gtrsim$~10--24 the heating rate and maximum temperature are largest and approximately independent of $t_p/t_A$, leading to a strong emission in X-rays and EUV. On the opposite decreasing $t_p/t_A$  to smaller values leads to lower heating rates and plasma temperatures, and consequently fading high-energy radiative emission once $t_p/t_A \lesssim$~1--3. The average volumetric loop heating rate is shown to scale as $\ell_p u_p B_0^2/4\pi L^2$, where $\ell_p$ and $u_p$ are respectively the convective granule length-scale and velocity, $B_0$ is the intensity of the strong magnetic field threading the loop, and $L$ the loop length. These findings support a recent hypothesis explaining ultracool dwarf observations of stars with similar magnetic field strength but radically different topologies displaying different radiative emission. 
\end{abstract}




\section{Introduction}

The high-energy emission (X-rays and EUV) from late-type main sequence
stars is sustained, as in the solar case, by surface convective motions and a dynamo mechanism that combined give rise to million degree magnetized
coronae and stellar winds \citep{2004A&ARv..12...71G, 2006SoPh..234...41K, 2015RSPTA.37340259T}.

Bright X-ray solar coronal loops are arch-like structures rooted in the photosphere and threaded by a strong magnetic field, with their field lines connecting photospheric regions of opposite polarity at their footpoints. Because typical loop lengths $L$ are about 10--100$\times 10^3$\,km while the Alfv\'en speed $v_A$ is about $2 \times 10^3$\,km/s \citep{2014LRSP...11....4R}, the resulting Alfv\'en crossing time is approximately given by $t_A=L/v_A \sim$~5--50\,s.
On the other hand the photospheric granule convective time-scale is in the range $t_p \sim$~ 5--10\,minutes  and consequently $t_p/t_A > 1$. Therefore the wavelength of the modes injected into the upper atmosphere by photospheric displacement of magnetic field line footpoints will be much longer than typical loop lengths, but it will have transverse scales of the order of the granulation correlation lengths, so that the wave will seem to be a perturbation whose wavelength is so long that it will appear as a magnetic field perturbation which is quasi-invariant along the loop \citep{Rappazzo2017}. Another aspect of the fact that the power injected by photospheric motions is slow, is that from the point of view of field resonances (see below) it is essentially equivalent to a zero frequency, i.e. quasi-constant, forcing. In the language of coronal heating theories this is called DC \citep{1991RvMA....4....1P}heating, with only a smaller contribution existing at frequencies beyond the first loop resonance, the so-called (AC) frequencies \citep{1997ApJ...490..442M}.

Though granulation injects energy in the upper layers at transverse scales of about $\sim 10^3$\,km \citep{2012ApJ...756L..27A}, smaller than typical loop lengths, the scales remain relatively large compared to the scales where energy dissipation is effective   in the high Lundquist number solar plasma (where $S=L v_A/\eta \sim 10^{13}$, and $\eta$ is the magnetic diffusivity). 
However, as originally proposed by \cite{1972ApJ...174..499P, 1988ApJ...330..474P, 1994ISAA....1.....P, 2012msdp.book....3P}, when the amplitude of the transverse magnetic field generated by footpoint motions grows beyond a certain threshold the magnetic field cannot remain in equilibrium at large scales \citep{2015ApJ...815....8R}: its relaxation involves the formation of small-scale current sheets on fast ideal timescales \citep{2013ApJ...773L...2R}, where magnetic energy can be efficiently dissipated.
The continuous shuffling of magnetic field line footpoints at the photosphere provides a source of magnetic energy which then dynamically evolves, leading to a statistically steady state where energy is efficiently transferred from the large injections scales to small scale structures (current sheets) where it can be efficiently dissipated and heat the loop plasma. As shown by direct \citep{1996ApJ...457L.113E, 1997ApJ...484L..83D, 1998ApJ...497..957G, 1998ApJ...505..974D, 1999ApJ...527L..63D, 2007ApJ...657L..47R, 2008ApJ...677.1348R, 2011PhRvE..83f5401R} and shell models \citep{2004PhRvL..92s4501N, 2005ApJ...629L.133N, Buchlin2007} numerical simulations this physical process can be  described as a particular instance of magnetically dominated MHD turbulence.
Even though line-tying keeps the system far from equipartition (with kinetic energy in the computational box much smaller than magnetic energy), a turbulent energy cascade with preferential energy transfer in the planes orthogonal to the strong axial magnetic field and a broad-band power-law magnetic energy spectrum develop. But the enhanced field line stiffness introduced by line-tying brings about steeper magnetic energy spectra $E_M(k_\perp) \propto k_{\perp}^{-\alpha}$ with $\alpha \in [5/3, 3]$, with the steepest spectra corresponding to stronger guide fields \citep{2003PhPl...10.3584D, 2007ApJ...657L..47R, 2008ApJ...677.1348R, 2011PhRvE..83f5401R}. Recent three-dimensional fully compressible simulations of the Parker model have shown that the UV emission computed for coronal loops with parameters (length, magnetic field intensity) typical of observed EUV-bright loops has properties compatible with observations \citep{2012A&A...544L..20D,2016ApJ...817...47D}.

However on the Sun there are also loops with lengths and magnetic field intensities such that the Alfv\'en crossing timescale is larger than the photospheric timescale, {\it i.e.,} $t_p/t_A < 1$. For these loops the low-frequency approximation used in the aforementioned works does not hold. Understanding the physical processes at work in these loops and their radiative emission can help to explain why some class of loops, e.g. large streamers loops, do not shine brightly in X-rays and EUV. In addition, stellar parameters (e.g., stellar mass, radius, rotational velocity, magnetic field intensity, photospheric temperature) change with stellar type, affecting the geometrical extent of closed structures in the corona, and also the magnetic topology, stellar granulation length- and time-scales, and possibly the dynamo mechanism.

In particular it is well known that for solar-like and later dwarfs magnetic field intensity and the ratio of X-ray to bolometric luminosity ($L_X/L_{bol}$) display similar behavior as a function of Rossby number $Ro=P_{rot}/\tau_c$ (where $P_{rot}$ is the rotational period and $\tau_c$ the convective turnover time). The average magnetic field intensity at first increases as a power-law toward smaller $Ro$ (higher rotational periods) until it saturates approximately at a few kilogauss for the fastest rotators with $Ro < 0.1$ \citep{2012LRSP....9....1R, Saar1996, Saar2001, Reiners2009, Reiners2010}. Normalized X-ray luminosity behaves similarly with saturation level $L_X/L_{bol} \approx 10^{-3}$ \citep{James2000, 2008ApJ...673.1080B, Reiners2010, 2011ApJ...743...48W}.

Solar-like and early M dwarfs follow these relationships well, with small deviations around the average. However these break down for so-called \emph{ultracool dwarfs} (UCDs), that are fast rotators later than about M7 with $Ro < 0.1$. Specifically, UCDs exhibit significantly more scatter  around their saturation value for $L_X/L_{bol}$ \citep{2012LRSP....9....1R}, with many of them displaying $L_X/L_{bol}$ as low as $\approx 10^{-5}$, a behavior possibly linked to a strong variation of the magnetic field topology \citep{2014ApJ...785...10C}.

Overall, different values of characteristic parameters, primarily through the ratio of the photospheric convective timescale over the Alfv\'en crossing time ($t_p/t_A$), can have a strong impact on the coronal loop dynamics and their investigation can advance our understanding of solar and stellar X-ray and EUV radiative emission. This motivates our investigation of the  heating of magnetically confined coronal regions due to field line footpoint shuffling for different values of $t_p/t_A$.

\section{Physical Model}

Our coronal models consider loops with large aspect ratios, with transverse scales smaller than the axial scale, which we treat in a simplified Cartesian model neglecting curvature: a loop is modeled as a ``straightened-out'' box with orthogonal square cross section of size $\ell$ (x-y planes) and axial length $L$ (z-direction).

\subsection{Governing Equations}

Numerical simulations are performed with two different sets of equations. Reduced magnetohydrodynamics (MHD) and fully compressible MHD. The reduced MHD model allows simulations with higher numerical resolutions and longer time durations, while with the more numerically demanding compressible MHD code simulations we are able to compute the temperature evolution and mean profiles in select cases. Here we discuss the reduced MHD model, while we defer the discussion on the compressible MHD model to Section~\ref{sec:cmhd}.

The reduced MHD approximation \citep{1974JETP...38..283K, 1976PhFl...19..134S, 1982PhST....2...83M, 1992JPlPh..48...85Z} describes well the dynamics of a plasma embedded in a strong axial homogeneous and uniform magnetic field $\mathbf{B}_0 = B_0 \mathbf{e}_z$, with a small ratio of kinetic to magnetic pressures, in the limit of a large loop aspect ratio and of a small ratio of poloidal to axial magnetic field intensities ($b/B_0 \le \ell/L \ll 1$). The density $\rho_0$ is then uniform and the velocity field $\mathbf{u}$ is incompressible and has no component in the axial direction $z$, in similar fashion to the fluctuating magnetic field $\mathbf{b}$ ($u_z=b_z=0$). Thus introducing the potentials $\varphi$ and $\psi$ we can write $\mathbf{u} = \nabla \varphi \times \mathbf{\hat{e}}_z$ and $\mathbf{b} = \nabla \psi \times \mathbf{\hat{e}}_z$, with the axial vorticity $\omega = -\nabla^2 \varphi$ and current density $j = -\nabla^2 \psi$. 

The reduced MHD equations in non-dimensional form are then given by
\begin{eqnarray}
	&&\partial_t \psi = [\varphi, \psi] + v_A \partial_z \varphi + \eta \nabla^2 \psi, \label{eq:rmhd1} \\
	&&\partial_t \omega = [j, \psi] -[\omega, \varphi] + v_A \partial_z j + \nu \nabla^2 \omega. \label{eq:rmhd2}
\end{eqnarray}
To render the equations non-dimensional the magnetic field has been expressed as an Alfv\'en velocity, i.e., $b \rightarrow b/\sqrt{4\pi\rho_0}$ (and for the guide field this has been indicated explicitly as $v_A =B_0/\sqrt{4\pi\rho_0}$), and then all velocities normalized to the photospheric granulation velocity rms $u_{p}$ ( $\sim 1$\,km/s for the Sun). The Poisson bracket is defined as, e.g., $[\varphi, \psi] = \partial_x \varphi \partial_y \psi - \partial_y \varphi \partial_x \psi$, and the Laplacian operator has only orthogonal components $\nabla = \mathbf{\hat{e}}_x \partial_x + \mathbf{\hat{e}}_y \partial_y$. Lengths are normalized to the orthogonal box length, thus $\ell$=1 and $L$=10 (because the aspect ratio $\ell/L$ is fixed to 10 in our simulations). Normalized resistivity and viscosity coefficients are set equal to $\eta$=$\nu$=1/$R$, where the Reynolds number $R$=800 and numerical resolution is $512^2 \times 200$.
More details on the numerical code and model are discussed in \cite{2007ApJ...657L..47R, 2008ApJ...677.1348R}.

\subsection{Boundary and Initial Conditions} \label{sec:bic}

As in previous simulations, line-tying boundary conditions are used at the bottom and top plates $z$=0 and $z$=L, i.e., the boundary condition imposed is a prescribed photospheric granulation-mimicking velocity field, with rms $u_p$, length scale $\ell_p$, and variation time scale $t_p$. In all our simulations we have chosen the linear box size to contain about 4 granules, i.e., $\ell_p \sim \ell / 4$. In the x--y planes a pseudo-spectral scheme with periodic boundary conditions is implemented, time is advanced with a third-order Runge-Kutta and an adaptive time step \citep[see][]{2007ApJ...657L..47R, 2008ApJ...677.1348R}.
Initially no velocity or fluctuating magnetic field are present in the computational box, i.e., $\mathbf{u} = \mathbf{b} = 0$ for $0 < z < L$.

In general photospheric motions characterized by a variation
timescale $t_p$ will not have a single harmonic at the frequency
$1/t_p$, rather the amplitude of its Fourier transform will peak
at the frequency  $1/t_p$ but include many other harmonics.
For this reason, as in previous work, we choose our boundary forcing velocity with velocity potential
\begin{eqnarray}
\varphi (x, y, t) = &&\frac{1}{\sqrt{\sum_{mn} \alpha^2_{mn}}}\, 
\sum_{k,l} \frac{\ell \alpha_{kl}(t)}{2\pi \sqrt{k^2+l^2}} \nonumber \\[.5em]
&&\sin \left[ \frac{2\pi}{\ell} \left( kx+ly \right)  + 2\pi \zeta_{kl}(t) \right]. \label{eq:forc1}
\end{eqnarray}
All wavenumbers $(k,l) \in Z^2$ included are in the range $3 \le (k^2+l^2)^{1/2} \le 4$ so that $\ell_p \sim \ell/4$.
The normalization in Eq.~(\ref{eq:forc1}) sets the velocity rms to $1/\sqrt{2}$.

The spatial pattern in Eq.~(\ref{eq:forc1}) changes with forcing timescale $t_p$. This is achieved via the coefficients $\alpha_{kl}(t)$ and $\zeta_{kl}(t)$, which are two sets of random numbers $\alpha_{kl}$ and $\zeta_{kl}$, each with values between 0 and 1,  independently chosen for the two boundary planes z=0 and z=L. At time $t=t_1$ we select two sets of random numbers $\alpha^1_{kl}$ and $\alpha^2_{kl}$ (and analogously $\zeta^1_{kl}$ and $\zeta^2_{kl}$), that when inserted in Eq.~(\ref{eq:forc1}) give rise to two different spatial patterns. The time dependent coefficients $\alpha_{kl}(t)$ and $\zeta_{kl}(t)$ in Eq.~(\ref{eq:forc1}) are then built in the following way
\begin{equation}
\alpha_{kl}(t) = \alpha^1_{kl}  + \frac{\alpha^2_{kl} - \alpha^1_{kl}}{t_p} \left( t - t_1 \right), \qquad \textrm{for} \qquad t \in [t_1, t_1+t_p]   \label{eq:forc2}
\end{equation}
(and analogously for $\zeta_{kl}(t)$). In this way $\alpha_{k,l}(t)$ goes from  $\alpha^1_{kl}$ at time $t=t_1$ to $\alpha^2_{kl}$ at time $t=t_1+t_p$. This procedure is then repeated for consecutive time intervals of size $t_p$.

\subsection{Dimensional Analysis} \label{sec:da}

Photospheric granular cells of length $\ell_p$ and characteristic velocity $u_p$ can be dimensionally associated with the turnover timescale $\tau_p = \ell_p/u_p$. In the solar case the photospheric velocity variation timescale $t_p$, i.e.,  the average lifetime of a granular convective cell, has typical value $t_p \sim$~5--10 minutes, and is approximately of the same order as the turnover time $t_p \sim \tau_p$. Here we will assume that the relation $t_p \sim \tau_p = \ell_p/u_p$ is approximately valid in general for stars with a convective envelope, and we use it as a baseline for our calculations. Nevertheless as discussed in more detail in Section~\ref{sec:ns} we also perform simulations with values of $t_p$ different from $\tau_p$, in order to change the value of $t_p/t_A$, and understand how this affects the dynamics.

The dimensional analysis of Eqs.~(\ref{eq:rmhd1})-(\ref{eq:rmhd2}) shows that when $t_p\sim \tau_p = \ell_p/u_p$, apart from the Reynolds numbers, its solutions are a one-parameter family depending only by one fundamental parameter:
\begin{equation} \label{eq:rmhdp1}
f_{pA} = \frac{t_p}{t_A} = \frac{v_A \ell_p}{u_p L},
\end{equation}
that is the ratio of the photospheric convective timescale $t_p \sim \tau_p = \ell_p/u_p$ over the axial Alfv\'en crossing time $t_A = L/v_A$.

This conclusion can be inferred from the non-dimensional equations (\ref{eq:rmhd1})-(\ref{eq:rmhd2}), where we have used the box orthogonal length $\ell$ and the photospheric velocity $u_p$ as characteristic quantities to render the equations non-dimensional. Because the box length $\ell$ is fixed at about 4 granule length-scales $\ell \sim 4 \ell_p$ (see Section~\ref{sec:bic}) we are effectively using the relevant physical length $\ell_p$, velocity $u_p$, and associated timescale $\tau_p=\ell_p/u_p$ to render the equations non-dimensional.

Therefore the only coefficient present in Eqs.~(\ref{eq:rmhd1})-(\ref{eq:rmhd2}) in front of the linear terms $\propto \partial_z$ is $v_A$ in non-dimensional units, i.e., $v_A/u_p$, where $v_A$ is the Alfv\'en velocity associated to the axial magnetic field ($v_A = B_0/\sqrt{4\pi\rho_0}$). Furthermore, in order to render explicit the dependence of the solutions by the axial box length $L$, we can make the change of variable $z' = z\, \ell_p/L$ with $z' \in [0,1]$. Thus finally we obtain in front of the linear terms the coefficient $f_{pA} = v_A\ell_p/(u_pL)$ (Eq.~(\ref{eq:rmhdp1})), with a computational box of unitary size along all three spatial directions. 

Hence, all non-dimensional solutions of Eqs.~(\ref{eq:rmhd1})-(\ref{eq:rmhd2}) depend on this single parameter $f_{pA}$ \citep[e.g., see][]{2008ApJ...677.1348R}. Therefore once we find the non-dimensional solution for a given value of $f_{pA}$ it will be the same solution for different loops with different values of axial length $L$, Alfv\'en velocity $v_A$, granulation velocity $u_p$ and granulation length-scale $\ell_p$ but with same $f_{pA}$. Related physical quantities, such as energies, Poynting flux, and heating rate, can then be derived from the fields $\mathbf{u}$ and $\mathbf{b}$.

Although the non-dimensional solutions are the same for a given value of $f_{pA}$, the corresponding dimensional solutions will in general be different, since the quantities used to render the equations non-dimensional will in general change. Explicitly, indicating dimensional quantities with the subscript `dim', the dimensional velocity and magnetic fields are obtained multiplying for the dimensional factor $u_p$, i.e., $\mathbf{u}_{dim} = u_p \mathbf{u}$ and $\mathbf{b}_{dim} = u_p \sqrt{4\pi\rho_0}\, \mathbf{b}$. The Poynting flux is a quantity of strong interest because it gives the energy per unit time and surface entering the computational box from the photosphere-mimicking boundary planes z=0 and L. The non-dimensional Poynting flux along z (i.e., along the direction normal to the photospheric boundaries) is given by $S_z=-v_A \mathbf{u} \cdot \mathbf{b}$ \citep[e.g., see Section 3.1 in][]{2008ApJ...677.1348R}, and to obtain the dimensional Poynting flux we need to multiply for the dimensional factor $\rho_0 u_p^3$, i.e., $S_{z, dim} = -\rho_0 u_p^3\, S_z$, a quantity that clearly in dimensional units is not dependent only on the ratio $f_{pA}$. In fact $\rho_0$, $u_p$, and $v_A$ appear explicitly and cannot be reduced to $f_{pA}$. Therefore the Poynting flux depends also directly on these quantities, as well as on $f_{pA}$ through the non-dimensional velocity and magnetic fields $\mathbf{u}$ and $\mathbf{b}$ that do depend exclusively by $f_{pA}$.

Notice that $f_{pA}$ can also be written as the fractional wavelength of the Alfv\'en wave propagating from the boundary along the axial direction $z$ associated with the granular motion timescale $t_p$, i.e.,
\begin{equation} \label{eq:rmhdp2}
f_{pA} = \frac{t_p}{t_A} = \frac{\lambda_\parallel}{L},
\end{equation}
the ratio of the parallel wavelength $\lambda_\parallel$ over the axial loop length $L$.
Solar X-ray and EUV bright loops typically have $t_p/t_A > 1$ ($\sim$~10--100),
for which the wavelength associated to the boundary forcing is larger than the loop length, so that the system has a strong two-dimensional component.
Loops with  $t_p/t_A < 1$ have a smaller or negligible high-energy radiative emission (for instance larger solar streamer loops do not exhibit significant X-ray emission), and wavelength smaller than the loop length.
In previous works we have always considered loops with $t_p/t_A \gg 1$ \cite[e.g.][]{2007ApJ...657L..47R, 2008ApJ...677.1348R}, and used the so-called \emph{zero-frequency approximation}, i.e., kept the boundary velocity forcing constant in time corresponding to $t_p/t_A \rightarrow \infty$ instead of using a slow-varying velocity. In Section~\ref{sec:ns} we first of all check that a photospheric velocity which is constant in time is a good approximation for large values of $t_p/t_A$, and then investigate the dynamics also for smaller values of $t_p/t_A$ to understand the origin of the diminished radiative emission for these kind of loops and improve our understanding of X-ray and EUV emission from the Sun and other stars.

\subsection{Linear Analysis} \label{sec:la}

A first insight into the coronal dynamics for different values of $t_p$ comes from a linear analysis of wave propagation in these loops, which we briefly summarize here. Introducing the Els\"asser variables $\mathbf{z}^{\pm} = \mathbf{u} \pm \mathbf{b}$ and neglecting nonlinear and diffusive terms in Equations~(\ref{eq:rmhd1})-(\ref{eq:rmhd2}), the remaining linear terms yield the two wave equations:
\begin{equation}
	\partial_t \mathbf{z}^\pm = \pm v_A\, \partial_z \mathbf{z}^\pm, \qquad \textrm{with} \qquad \nabla \cdot \mathbf{z}^\pm = 0.
\end{equation}
Thus, fluctuations propagate along $z$ at the Alfv\'en speed $v_A$. Line-tying boundary conditions imply a ``reflection'' at the boundary. For instance, a velocity $\mathbf{u}_p$ at the boundary z=0 implies the condition $\mathbf{z}^- = -\mathbf{z}^+ + 2 \mathbf{u}_p$ at that boundary ($\mathbf{z}^-$ propagates inward, $\mathbf{z}^+$ outward), i.e., generalized Alfv\'en waves are continuously injected and propagate in the computational box. 

Considering an initial condition with only the axial guide field with $\mathbf{v}_A = v_A \mathbf{\hat{e}}_z$, vanishing orthogonal component $\mathbf{b} = 0$, and a forcing velocity with frequency $\nu$  at the photospheric boundary that shuffles the magnetic field line footpoints, i.e., 
\begin{equation} \label{eq:vres}
\mathbf{u}(x,y,z=L,t) = \mathbf{u}_{p} (x,y) \cos(\omega t),
\end{equation}
where $\omega = 2\pi \nu$ is the angular frequency, this 
will inject Alfv\'en waves at that frequency propagating in the axial direction of the loop.

In general these waves, that are continuously injected and reflected at the boundaries, will be out of phase and decorrelated along
the loop so that their sum will remain limited the whole time to values of the order
of the forcing velocity at the boundary, with no  growth in time for the amplitude
of the resulting velocity and magnetic fields.
But for the \emph{resonant frequencies} 
\begin{equation} \label{eq:rf}
\nu_n =n\, \frac{\nu_A}{2}, \quad \textrm{with} \quad n \in \mathbb{N},\ \nu_A = \frac{1}{t_A},
\end{equation} 
at corresponding timescale $t_n = 2 t_A/n$, the waves are in phase and they sum coherently \citep{1985A&A...146..199I}, with the magnetic and velocity fields in the loop growing \emph{linearly in time}, i.e., the loops behave as resonant cavities. 
Using Eq.~(\ref{eq:vres}) with resonant frequency $\nu_n$ and $n \ge 1$, the resulting fields grow approximately as \citep{1999PhPl....6.4146E, 2010PhDT.......193R, 2015bps..book.....C}:
\begin{align}
&\mathbf{b} \sim \mathbf{u}_{p} \cos \left(\omega_n \frac{z}{v_A} \right) \cos( \omega_n t)\,  
\frac{t}{t_A}, \label{eq:linres1} \\
&\mathbf{u} \sim \mathbf{u}_{p} \sin \left( \omega_n \frac{z}{v_A} \right) \sin( \omega_n t)\, 
\frac{t}{t_A}. \label{eq:linres2}
\end{align}

As shown in subsequent sections, the \emph{zero frequency resonance} $n=0$ of the system, i.e., a boundary velocity $\mathbf{u}_p$ constant it time, is of fundamental importance for solar and stellar coronal heating.  This differs from resonances with $n \ge 1$ because while the magnetic field grows linearly in time, and approximately uniformly along~z
\begin{equation}
\mathbf{b} \sim \mathbf{u}_{p}\, \frac{t}{t_A}, \label{eq:lincst1}
\end{equation}
for $n=0$ the \emph{velocity field does not grow} and its value remains of the same order of magnitude of the photospheric velocity ($u \sim u_{p}$).

The boundary velocity Eq.~(\ref{eq:forc1}) does not have a single mode as in Eq.~(\ref{eq:vres}) but will in general result in a superposition of modes peaked at frequency $1/t_p$.
As mentioned in Section~\ref{sec:da}, for X-ray bright solar coronal
loops photospheric motions have a low frequency ($t_p/t_A >1$),
giving rise to a coronal magnetic field dominated by 2D modes invariant along $z$,
and their low frequency can then be approximated with zero as shown in the following Section~\ref{sec:ns}

The linear analysis shows that magnetic energy buildup in the corona only results from resonant forcing frequencies. Notice though that since $t_n/t_A = 2/n$, resonance timescales $t_n$ for $n \ge 2$ are tightly packed in the interval $[0,t_A]$. Therefore we can expect that our forcing velocity at timescale $t_p$ (Eq.~(\ref{eq:forc1})) for $t_p \le t_A$ will always have some component at a resonance frequency.

\section{Numerical Simulations} \label{sec:ns}

In this section we present a series of numerical simulations modeling a coronal layer driven by a boundary forcing velocity with different frequencies (Eqs.~(\ref{eq:forc1})-(\ref{eq:forc2})). On the bottom and top planes we employ two independent velocity forcings with characteristic timescale $t_p$ as described in Section~\ref{sec:bic}. As discussed in Section~\ref{sec:da} we make the \emph{ansatz} that the timescale of convective cells $t_p$ (their lifetime) is in general given by the granular turnover time $\tau_p =\ell_p/u_p$, i.e., $t_p \sim \tau_p=\ell_p/u_p$. This relations works for solar convection, and we assume it to work also for convective motions in other stars, but clearly further observational measurements to confirm or modify this assumption would be helpful. 

As discussed in Section~\ref{sec:da} the non-dimensional reduced MHD solutions for $\mathbf{u}$ and $\mathbf{b}$ depend only by the parameter $f_{pA}=t_p/t_A$ when the forcing frequency is set to the granular turnover time ($t_p \sim \tau_p = \ell_p/u_p$). At first we want to test the validity of the \emph{zero-frequency approximation}, i.e., if a velocity constant in time is a good proxy for a slowly changing one, namely if for loops with $t_p/t_A > 1$ we can use a vanishing forcing frequency in place of a low one.
Additionally we also investigate the effect of a photospheric velocity that changes on shorter timescales all the way down to the Alfv\'en crossing time $t_A$ or shorter. To this effect we change $t_p$ independently of $\tau_p = \ell_p/u_p$. Namely, in the following simulations we keep every other parameter fixed, including $L$, $v_A$ (and therefore $t_A$),  $\ell_p \sim \ell/4$ and $u_p = 1/\sqrt{2}$, but change the velocity timescale $t_p$ (Eqs.~(\ref{eq:forc1})-(\ref{eq:forc2})). This approach is numerically less demanding, and therefore more feasible, than changing the loop length or Alfv\'en velocity. Additionally, for numerical reasons, we use a value of $v_A=200$, corresponding to an Alfv\'en velocity of $200$\,km/s, smaller than the observationally estimated value of $2000$\,km/s for the Sun. 

For typical Solar X-ray bright loops $t_p/t_A > 10$. In fact since $t_p \sim$~5--10\,minutes (typical solar granulation timescales) and observationally estimated values for typical Alfv\'en velocity is $v_A$=2,000\,km/s and loop length $L$=40,000\,km, we then obtain $t_A = 20$\, s and therefore an estimated value of $t_p/t_A \sim$~15--30.

\begin{figure}
	\includegraphics[width=\columnwidth]{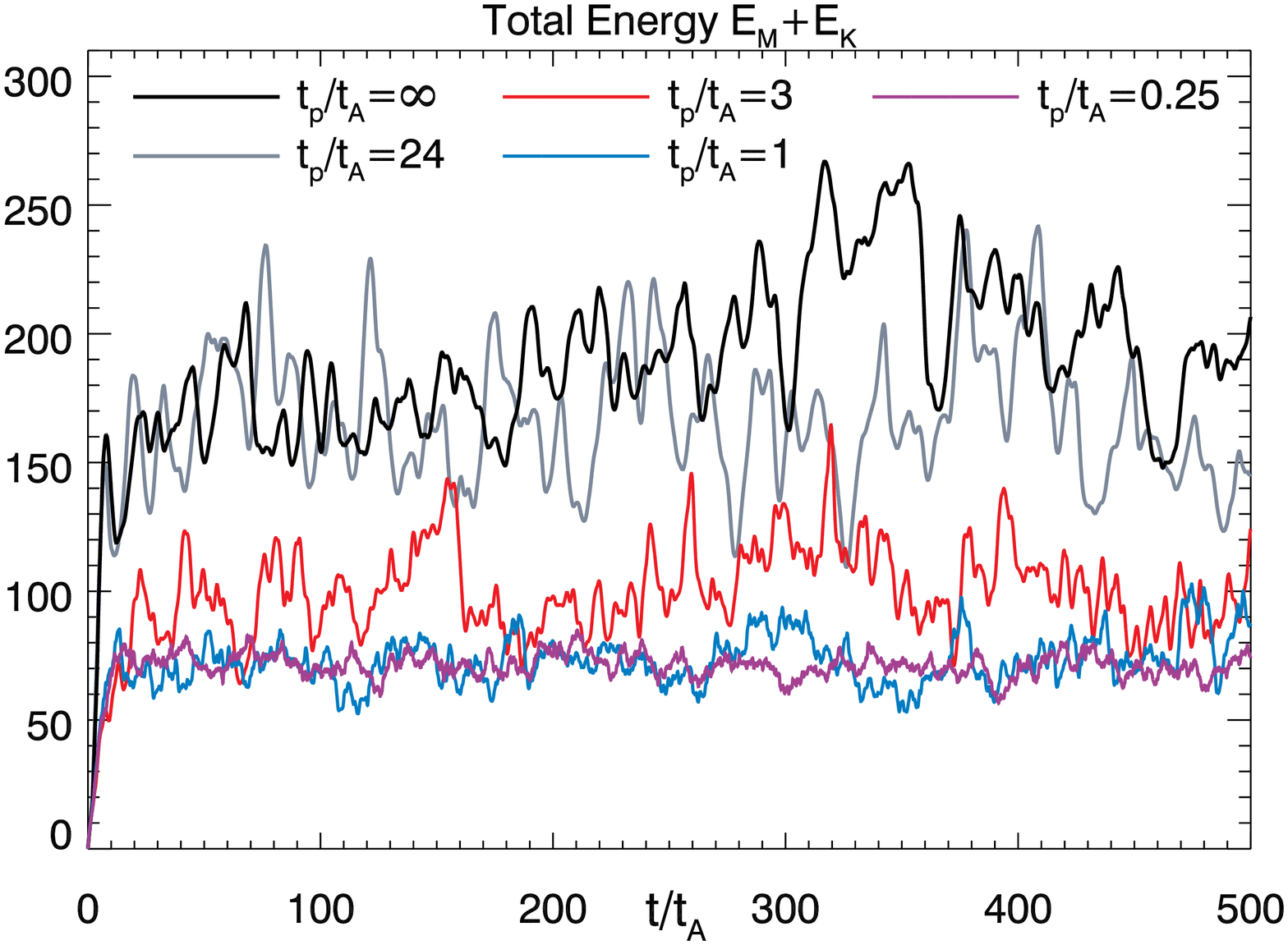}\\[1em]
	\includegraphics[width=\columnwidth]{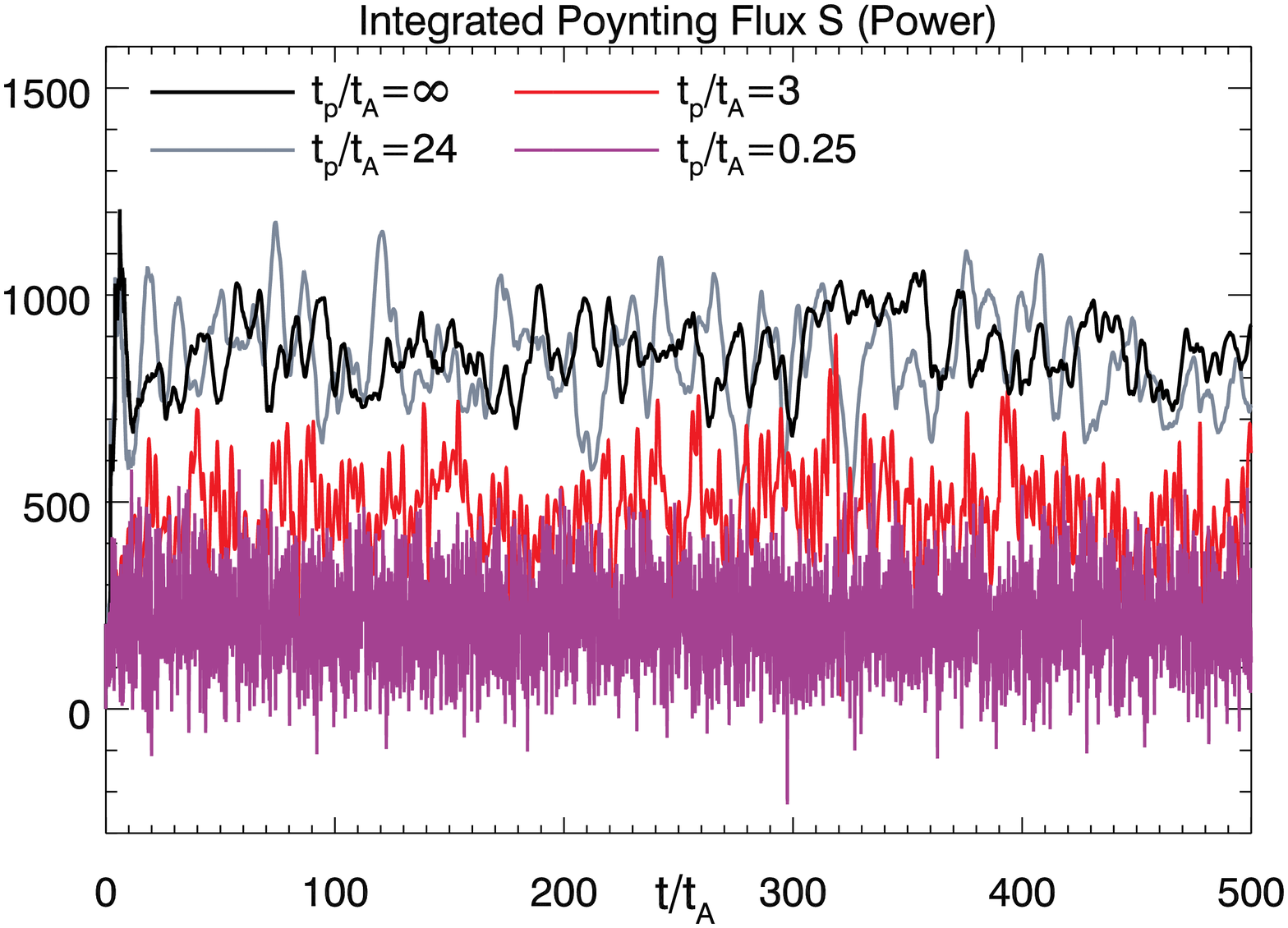}\\[1em]
	\includegraphics[width=\columnwidth]{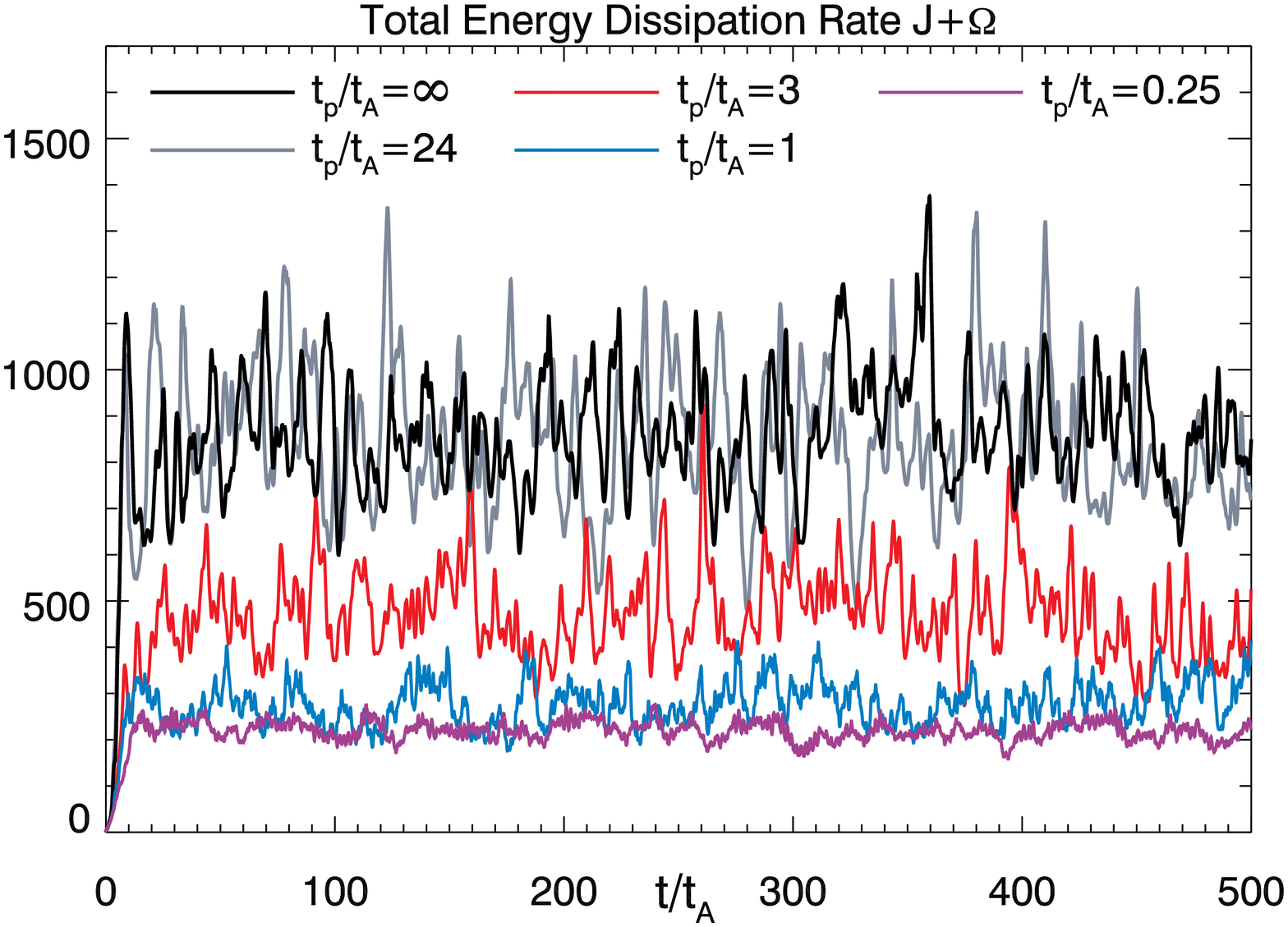}
	\caption{Total (magnetic + kinetic) energy versus time (normalized to the Alfv\'enic crossing time $t_A=L/B_0$) for simulations with different values of $t_p/t_A$ are shown in the \emph{top panel}. In similar fashion the \emph{middle panel} shows the time evolution of the integrated Poynting flux, and the \emph{bottom panel} for the total (ohmic plus viscous) dissipation rate. For all values of $t_p/t_A$ the system reaches in the nonlinear stage a statistically steady state.}
	\label{fig:en}
\end{figure}

\begin{figure*}
	\begin{center}
		\includegraphics[width=.48\textwidth]{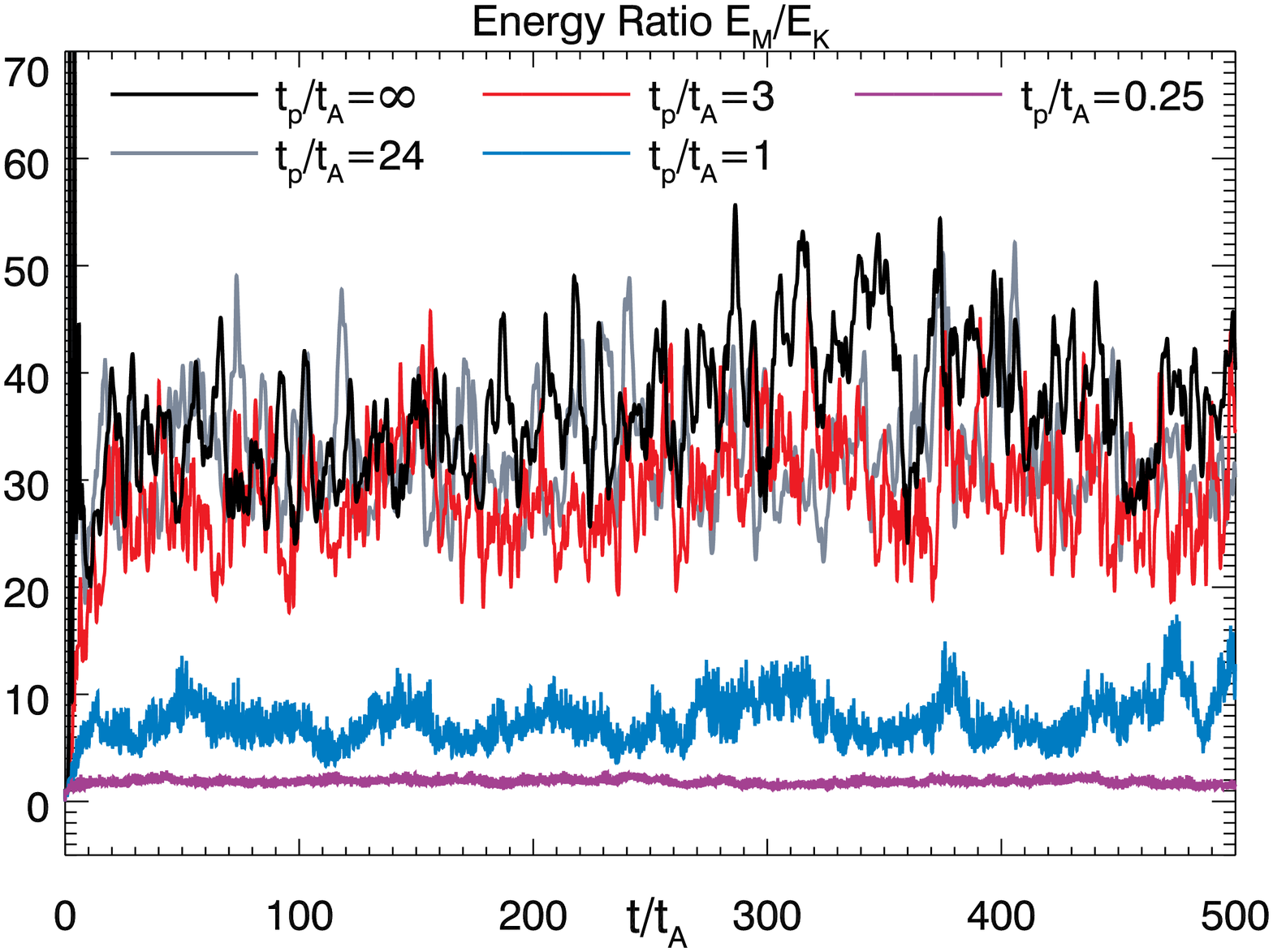} \hspace{1em}
		\includegraphics[width=.48\textwidth]{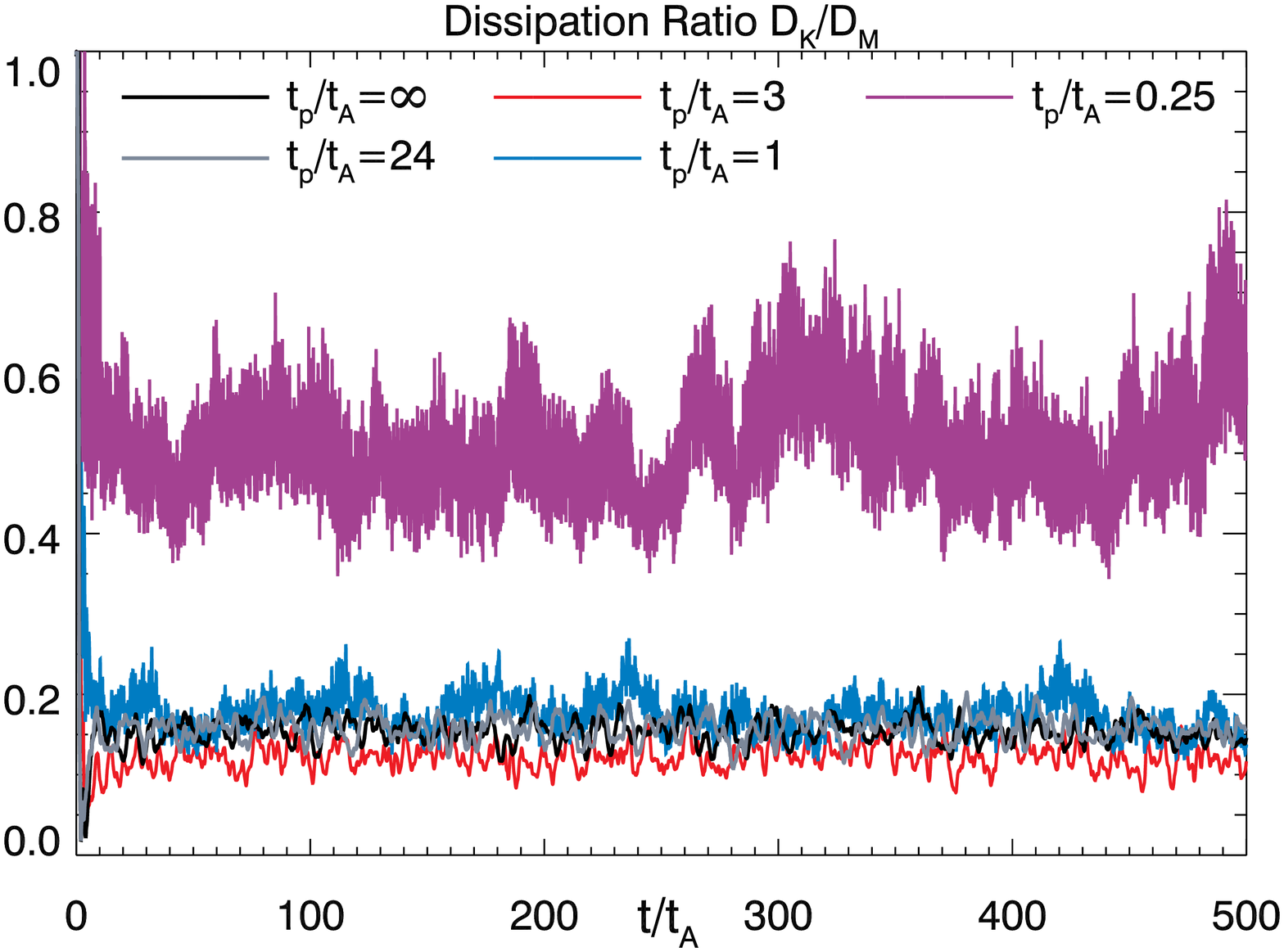}
	\end{center}
	\caption{The evolution in time of the ratio between magnetic ($E_M$) and kinetic ($E_K$) energy is shown in the \emph{left panel}, while in the \emph{right panel} that of the ratio of viscous ($D_K$) over ohmic ($D_M$) dissipation rates is shown.}
	\label{fig:enr}
\end{figure*}

In previous work \citep{2007ApJ...657L..47R, 2008ApJ...677.1348R} we have considered boundary photospheric motions \emph{constant} in time, the so-called zero-frequency approximation. Indeed for X-ray bright solar loops with $t_p/t_A > 10$ we expect that there should be little difference between a slowly varying photospheric velocity pattern and a constant one. In Figure~\ref{fig:en} the simulation with constant photospheric motions is indicated with $t_p/t_A = \infty$ to imply that the boundary forcing does not change in time. As shown in Figures~\ref{fig:en} and \ref{fig:enr} \emph{the dynamics are very similar} for the runs with $t_p/t_A = 24$ and $t_p/t_A = \infty$. As in our previous work the dynamics are initially well-described by the analysis of the zero-frequency resonance discussed in Section~\ref{sec:la}, with the velocity field remaining small and of the order of its boundary value $\mathbf{u} \sim \mathbf{u}_p$, while the magnetic field $\mathbf{b}$ grows linearly in time, implying a quadratic growth for the energy. Here magnetic ($E_M$) and kinetic ($E_K$) energies are defined respectively as:
\begin{equation} \label{eq:nrg}
E_M = \frac{1}{2} \int_V \mathrm{d}^3 \mathbf{x}\ \mathbf{b}^2, \qquad \textrm{and} \qquad
E_K = \frac{1}{2} \int_V \mathrm{d}^3 \mathbf{x}\ \mathbf{u}^2,
\end{equation}
and total energy as their sum $E = E_M + E_K$.
Once the magnetic field becomes large enough the system transitions to the nonlinear stage, and turbulent nonlinear dynamics sets in, leading to a statistically steady state with total energy fluctuating around a mean value (Figure~\ref{fig:en}, top panel). As shown in Figure~\ref{fig:enr} (left panel) the energy ratio $E_M/E_K$ in the nonlinear stage has similar averages for the two simulations with $E_M/E_K \sim 37$ for $t_p/t_A = \infty$ and $E_M/E_K \sim 33$ for $t_p/t_A = 24$.

The work done by photospheric motions on magnetic field lines footpoints in the planes $z$=0 and $z$=L injects energy in the computational box. This energy flux is quantified by the axial component of the Poynting flux \citep{2008ApJ...677.1348R}, given by
\begin{equation} \label{eq:poy}
	S_z = \mathbf{S} \cdot \mathbf{\hat{e}}_z = -v_A\ \mathbf{u_p} \cdot \mathbf{b}.
\end{equation}
The total energy entering the computational box per unit time is given by the surface integral of the Poynting flux:
\begin{equation} \label{eq:poyint}
S = v_A \left( \int_{z=L} da\ \mathbf{u_p} \cdot \mathbf{b} - \int_{z=0} da\ \mathbf{u_p} \cdot \mathbf{b} \right),
\end{equation}
with $\mathbf{u_p}$ prescribed by boundary conditions in $z$=0 and L. The contribution of the orthogonal components of the Poynting flux, and of the remaining box surfaces, to the energy equation is null because of periodicity in $x$-$y$ planes.
Indicating ohmic and viscous dissipations respectively with
\begin{equation}
D_M = \eta \int_V \mathrm{d}^3 \mathbf{x}\ j^2, \qquad
D_K = \nu \int_V \mathrm{d}^3 \mathbf{x}\ \omega^2,
\end{equation}
the energy equation for the system is given by
\begin{equation}
	\frac{\partial E}{\partial t} = S - (D_M + D_K).
\end{equation}
Therefore the energy balance of the system is due to the competition between the energy (Poynting flux) flowing into the computational box and the ohmic and viscous dissipation \citep[for a more detailed discussion, see section 3.1 in][]{2008ApJ...677.1348R}. 
As shown in Figure~\ref{fig:en} (middle and bottom plates) the integrated Poynting flux and total energy dissipation rate also fluctuate around their common mean value balancing each other, with their average about the same for both simulations with $t_p/t_A = \infty$ and 24, with respectively $\langle S \rangle = \langle D_M+D_K \rangle \sim$ 864 and 842.
Additionally the preponderance of magnetic over kinetic energy in these two simulations (Figure~\ref{fig:enr}, left panel, $E_K/E_M \sim 0.03$) is reflected also in the higher value of ohmic over viscous dissipation, with $D_K/D_M \sim 0.15$ for both simulations, as shown in the right panel of Figure~\ref{fig:enr}.

The two simulations that we have just analyzed show that for typical solar X-ray bright loops (with $t_p/t_A \gtrsim 10$) we can approximate the slow-changing boundary velocity as constant in time because there is little difference in the dynamics and heating rates.  Using this zero-frequency approximation \cite{2007ApJ...657L..47R, 2008ApJ...677.1348R} had derived scaling laws for the Poynting flux and heating rates as a function of loop parameters, showing that they are compatible with the observational constraint (further confirmed by the more recent fully compressible radiative simulation by \cite{2012A&A...544L..20D, 2016ApJ...817...47D}). As discussed in Section~\ref{sec:da} the non-dimensional solutions $\mathbf{u}$ and $\mathbf{b}$ of the reduced MHD equations depends only on the parameter $f_{pA} = v_A \ell_p/(u_p L)$ (Eq.~(\ref{eq:rmhdp1})). The scaling laws derived previously in the zero-frequency approximation are then still valid also for stars other than the Sun with different photospheric convective scale $\ell_p$ and velocity $u_p$, loop length $L$ and Alfv\'en velocity $v_A$, as long as $f_{pA}=t_p/t_A \gtrsim 10$. In the strong magnetic guide field limit (the most relevant case for most loops of interest) the Poynting flux $S_z$ in dimensional units is then given by \citep[see Eq.~(68) in][]{2008ApJ...677.1348R}:
\begin{equation}
S_z = \rho_0 v_A u_p^2 \left( \frac{v_A \ell_p}{u_p L}  \right)^\beta \ \textrm{erg cm}^{-2} \textrm{ s}^{-1}, \label{eq:pfstar}
\end{equation}
with $\beta =1$ in the strong guide magnetic field case \citep[see][for a more complete discussion of this scaling law and the value of $\beta$]{2008ApJ...677.1348R}.
A quantity of strong interest is the average volumetric heating rate, given by $S_z/L$ (because dissipation and energy injection rates are the same on average for this system in the statistically steady regime, see middle and bottom panels in Figure~\ref{fig:en} and \cite{2008ApJ...677.1348R}), that with $v_A=B_0/\sqrt{4\pi\rho_0}$ can be written as:
\begin{equation} \label{eq:vhstar}
\frac{S_z}{L} \sim  \frac{\ell_p u_p}{4\pi} \left( \frac{B_0}{L} \right)^2 \ \textrm{erg cm}^{-3} \textrm{ s}^{-1},
\end{equation}
using Gauss-cgs units. The average volumetric heating rate does not depend explicitly on the average density $\rho_0$, it is directly proportional to the convective granules velocity $u_p$ and lengthscale $\ell_p$, and depends quadratically on the ratio of the magnetic field intensity $B_0$ over the loop length $L$. 
The average volumetric heating rate outlined in Eq.~(\ref{eq:vhstar}) has been confirmed also by analytical calculations based on shell models by  \cite{2008ApJ...685..606N} (as long as the forcing velocity auto-correlation timescale is taken equal to the eddy turnover time $\tau_p = \ell_p/u_p$, that is an appropriate hypothesis). Additionally \cite{Malara2010} has extended this shell model calculation to inhomogeneous loops with gravitational density stratification along their axes, finding that the scaling~(\ref{eq:vhstar}) remains a good approximation.

Next we consider the case of loops with smaller values of $t_p/t_A$ for which we do not expect the zero-frequency approximation to be valid. Numerically it is very demanding to increase the value of the loop length $L$ or Alfv\'en velocity $v_A$ (and therefore $t_A$). Alternatively, since we are interested in understanding the effects of a boundary velocity that changes on timescales of the order of the Alfv\'en crossing time, our simulations keep all parameters same as in the previous simulations, changing only the value of the forcing frequency $t_p$ in Eqs.~(\ref{eq:forc1})-(\ref{eq:forc2})
\footnote{The forcing timescale $t_p$ in the case considered here is different from the granule turnover time $\tau_p \sim \ell_p/u_p$. We discuss further this point later in this section}.

We start with $t_p/t_A=3$. Resonant timescales are given by $\tau_n = 2t_A/n$ with $n=0,1,2,\ldots$, therefore $t_p/t_A=3$ is not a resonance but is significantly  smaller than $t_p/t_A \sim$~10--24, and close to the resonance with n=1. As shown in Figure~\ref{fig:en} energy, integrated Poynting flux and total dissipation rates are all significantly smaller respect to the runs with  $t_p/t_A=24$ and $\infty$. Nevertheless as can be seen in Figure~\ref{fig:enr} the magnetic energy is still larger than kinetic energy with $E_M/E_K \sim 29$, and also ohmic dissipation is larger than viscous dissipation with $D_K/D_M \sim 0.12$. Therefore the structure of the fields is still similar to the runs with $t_p/t_A=24$ and $\infty$, with a magnetic field more intense than velocity, but energy, dissipation and Poynting flux decrease substantially.

The runs with $t_p/t_A=1$ and $0.25$ have associated wavelengths (Eq.~(\ref{eq:rmhdp2})) smaller than the loop length since $\lambda_\parallel = t_p/t_A$. Additionally these timescales are also resonances. Figure~\ref{fig:en} shows that energy, Poynting flux and dissipation rates decrease further respect to simulations with higher $t_p/t_A$. Additionally Figure~\ref{fig:enr} shows that the structure of the fields is different. The magnetic over kinetic energy ratio is significantly smaller with $E_M/E_K \sim 7.6$ and $1.9$ respectively, while the viscous over ohmic dissipation ratio goes from $D_K/D_M \sim 0.18$ for $t_p/t_A=1$ to $D_K/D_M \sim 0.51$ for $t_p/t_A=0.25$. 

This implies that for the smaller values of $t_p/t_A$ the system approaches equipartition between kinetic and magnetic energies (and similarly for dissipation rates), akin to the linear fields for resonances with $n \ge 1$ (Equations~(\ref{eq:linres1})-(\ref{eq:linres2})).

\begin{figure}
	\includegraphics[width=.48\textwidth]{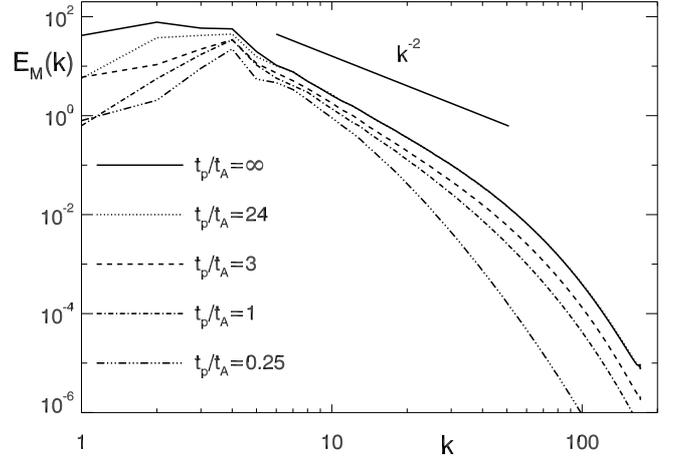}
	\caption{Magnetic energy spectra are shown for the simulations with different values of $t_p/t_A$. These values are time-averaged during the nonlinear stage of the dynamics.}
	\label{fig:sp}
\end{figure}

Further insight on the dynamics is given by the magnetic energy spectra $E_M(k)$ in perpendicular k-space (integrated along~$z$) shown in Figure~\ref{fig:sp}. The $t_p/t_A = \infty$ case (with same parameters as considered here) has been studied at high resolution showing that the magnetic energy spectrum exhibits an inertial range with a $k^{-2}$ power-law \citep{2011PhRvE..83f5401R}. Here the power-law index is smaller because of the lower numerical resolution and consequent heavier influence of diffusion. Nevertheless Figure~\ref{fig:sp} clearly shows that for all values of $t_p/t_A$ the system, that initially has energy only at large scales with wavenumbers k=3 and 4, develops a turbulent cascade that transfers energy toward the small scales where it is finally dissipated and the heating occurs. Although not shown in the Figure, magnetic and kinetic energy spectra confirm that for smaller values of $t_p/t_A$ the system approaches equipartition with the kinetic energy spectrum getting significantly closer to the magnetic one.
For lower values of $t_p/t_A$ the amount of energy at the small scales decreases significantly, implying that current sheets (that in the low frequency regime are the locations where energy dissipation occurs) tend to weaken and vanish at higher frequencies.
Because current sheets are the spatial structures where energy dissipation occurs, this is clearly fully in agreement with the corresponding decrease in total energy dissipation rates shown in Figure~\ref{fig:en} (bottom panel).

The main question to answer is therefore why the Poynting flux and consequently also the heating rates decrease from the zero-frequency case with $t_p/t_A \gtrsim$~10--24 toward smaller values of $t_p/t_A$. The Poynting flux decreases  also when $t_p$ is not at resonance, as for the case with $t_p/t_A=3$, and we can assume that it decreases incrementally also for all other values of   $3 \le t_p/t_A \le 10$. The reason is that the Poynting flux $S_z  = -v_A\ \mathbf{u_p} \cdot \mathbf{b}$ (Eqs.~(\ref{eq:poy})-(\ref{eq:poyint})) is determined by a \emph{scalar product}, and therefore the correlation, between the boundary velocity $\mathbf{u_p}$ and the magnetic field $\mathbf{b}$ at that boundary. 

As discussed in the linear analysis carried out in Section~\ref{sec:la}, it is only at the resonant frequencies that magnetic energy grows in time. 
This is a special case when the velocity and magnetic fields are in phase all the time, and the correlation between them at the boundary is maximized, so that a positive energy flux enters the computational box and the magnetic field can grow in time. For instance a detailed analysis of the solution with resonant velocity $\mathbf{u}^L(t) = \mathbf{u_p} \cos (\omega_n t)$ (Eq.~(\ref{eq:vres})) at the boundary z=L, shows that the magnetic field at that boundary is given by $\mathbf{b}^L(t) = \mathbf{u}^L(t)\, t/t_A$ so that the incoming Poynting flux from the boundary z=L is $S^L = v_A \mathbf{u}^L \cdot \mathbf{b}^L = v_A u_p^2 \cos^2(\omega_n t)\, t/t_A$. But if the special `resonant' conditions are not met exactly the fields at the boundary will decorrelate to an increasing extent the more the waves are out of phase, so that the energy flux entering into the box and the wave amplitudes' growth will progressively decrease.

On the other hand in the nonlinear stage the magnetic field is determined not only by the boundary forcing as in the linear stage, but also by the nonlinear dynamics developing in the numerical box. The turbulent dynamics will strongly modify the sinusoidal shape of the resonant waves, thus decorrelating the fields at the boundary and diminishing the Poynting flux. As shown by the numerical simulations the effect of the decorrelation in the nonlinear stage is stronger for higher resonance frequencies (higher $n$, Eq.~(\ref{eq:rf})). 

For low frequency of the boundary forcing with $t_p/t_A \gtrsim$~10--24 the magnetic field $\mathbf{b} \sim \mathbf{u}_{p}\, t/t_A$ is directly proportional to $\mathbf{u}_{p}$ (\ref{eq:lincst1}) thus maximizing the correlation between the two fields and the Poynting flux. Numerical simulations have shown that for the low-frequency case the correlation 
\begin{equation} \label{eq:corr}
C(z) = 
\frac{ \langle \mathbf{b}(z)\cdot\mathbf{u_p}(z) \rangle}
{ \langle \mathbf{b}^2(z) \rangle^{1/2} \langle \mathbf{u_p}^2(z) \rangle^{1/2} },
\end{equation}
where $\langle \ldots \rangle$ indicates the surface (x--y) average in a photospheric boundary (z=0 or L), decreases from 1 in the linear stage to about 0.65 in the nonlinear stage with a shear velocity boundary forcing \citep{2010ApJ...722...65R}, because turbulent dynamics modify the magnetic field particularly in the orthogonal x--y planes, but maintain a weak variation of the fields along z and a significant correlation with the velocity field at the boundary.

For $t_p/t_A = 3$ the wavelength of the waves injected at the boundary is about three box lengths $\lambda_{\parallel} \sim 3L$ (Eq.~(\ref{eq:rmhdp2})). Although it is longer than the box length the magnetic field has a significant variation over $\lambda_{\parallel}/3$, therefore a decreased decorrelation between the magnetic field and velocity at the boundary respect to $t_p/t_A \sim \lambda_{\parallel}/L \sim$~10--24 is expected, hence also a decreased Poynting flux as observed in the simulations.

The decorrelation between the magnetic and velocity fields at the boundary is further enhanced for smaller wavelength of the injected Alfv\'en waves, i.e., for smaller values of $t_p/t_A$. Indeed the correlation is necessary for the resonant fields to grow linearly in time during the initial stage (Eqs.~(\ref{eq:linres1})-(\ref{eq:linres2})), and same needs to occur for the Poynting flux. But in the nonlinear stage the specific correlation between magnetic and velocity field at the boundary is broken by the nonlinear dynamics that strongly decrease the correlation and consequently Poynting flux and heating rates as shown in Figure~(\ref{fig:en}).

\begin{figure}
	\includegraphics[width=.48\textwidth]{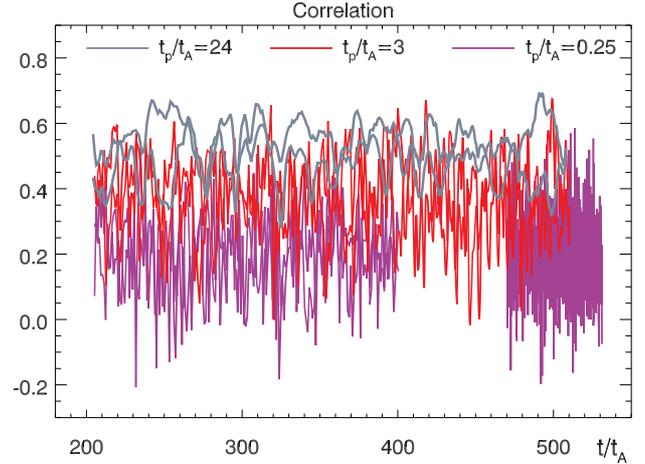}
	\caption{Correlation (Eq.~(\ref{eq:corr})) between magnetic and velocity fields at the boundaries z=0 and L as a function of time. We plot for each value of $t_p/t_A$ both $C(L)$ and $-C(0)$ that statistically have the same average. Correlation decreases for lower values of $t_p/t_A$, with time-averaged values equal to $\langle C \rangle =$~0.51, 0.38, and 0.20 for respectively $t_p/t_A =$~24, 3, and 0.25. For $t/t_A =$~0.25 there is a small gap in the data.}
	\label{fig:corr}
\end{figure}

The correlation C (Eq.~(\ref{eq:corr})) between magnetic and velocity fields at the boundaries in shown in Figure~\ref{fig:corr} for the simulations with $t_p/t_A$ = 24, 3, and 0.25, as a function of time in the approximate interval 200~$\lesssim t/t_A \lesssim$~540 (for $t/t_A =$~0.25 data are missing in a sub-interval). For an increased statistics we plot both $C(L)$ and $-C(0)$ for each simulation. The time average of these quantities are the same $\langle C(L) \rangle = - \langle C(0) \rangle$ for spatial statistical homogeneity, because the energy fluxes from both planes contribute equally in time (see Eq.~(\ref{eq:poyint})). The computed time averaged correlations confirm that it decreases for lower values of $t_p/t_A$, specifically $\langle C \rangle =$~0.51, 0.38, and 0.20 respectively for $t_p/t_A =$~24, 3, and 0.25, so that the correlation decreases more than half between $t_p/t_A =$~24 and 0.25. At the same time the integrated Poynting flux decreases from $S =$~834 to 219 (about 1/4), and magnetic energy from 161 to 46 (again about 1/4). This decreasing pattern for correlation, Poynting flux and magnetic energy is fully consistent with their expressions. In fact, indicating with $\ell^2$ the computational box cross-section surface, from Eq.~(\ref{eq:poyint}) we can estimate that $S \sim 2 v_A \ell^2 \langle \mathbf{u}_p \cdot \mathbf{b} \rangle$. Additionally the surface average $\langle \mathbf{b}^2 \rangle$ in Eq.~(\ref{eq:corr}) is approximately linked to magnetic energy $E_M$ (Eq.~\ref{eq:nrg}) as $\ell^2 \langle \mathbf{b}^2 \rangle \sim 2E_M/L$. Therefore because with our normalization $ \ell^2 \langle \mathbf{u_p}^2 \rangle = 1/2$, we can obtain from Eq.~(\ref{eq:corr}) that these three quantities are approximately linked as
\begin{equation}
     C \sim \frac{1}{2} \frac{\sqrt{L}}{v_A} \frac{S}{\sqrt{E_M}}.
\end{equation}
Consequently, since in our simulations $L$ and $v_A$ are fixed parameters, if the Poynting $S$ and magnetic energy $E_M$ both decrease as 1/4 then the correlation $C$ will decrease approximately by half.

\begin{figure*}
	\begin{center}
		\includegraphics[width=.48\textwidth]{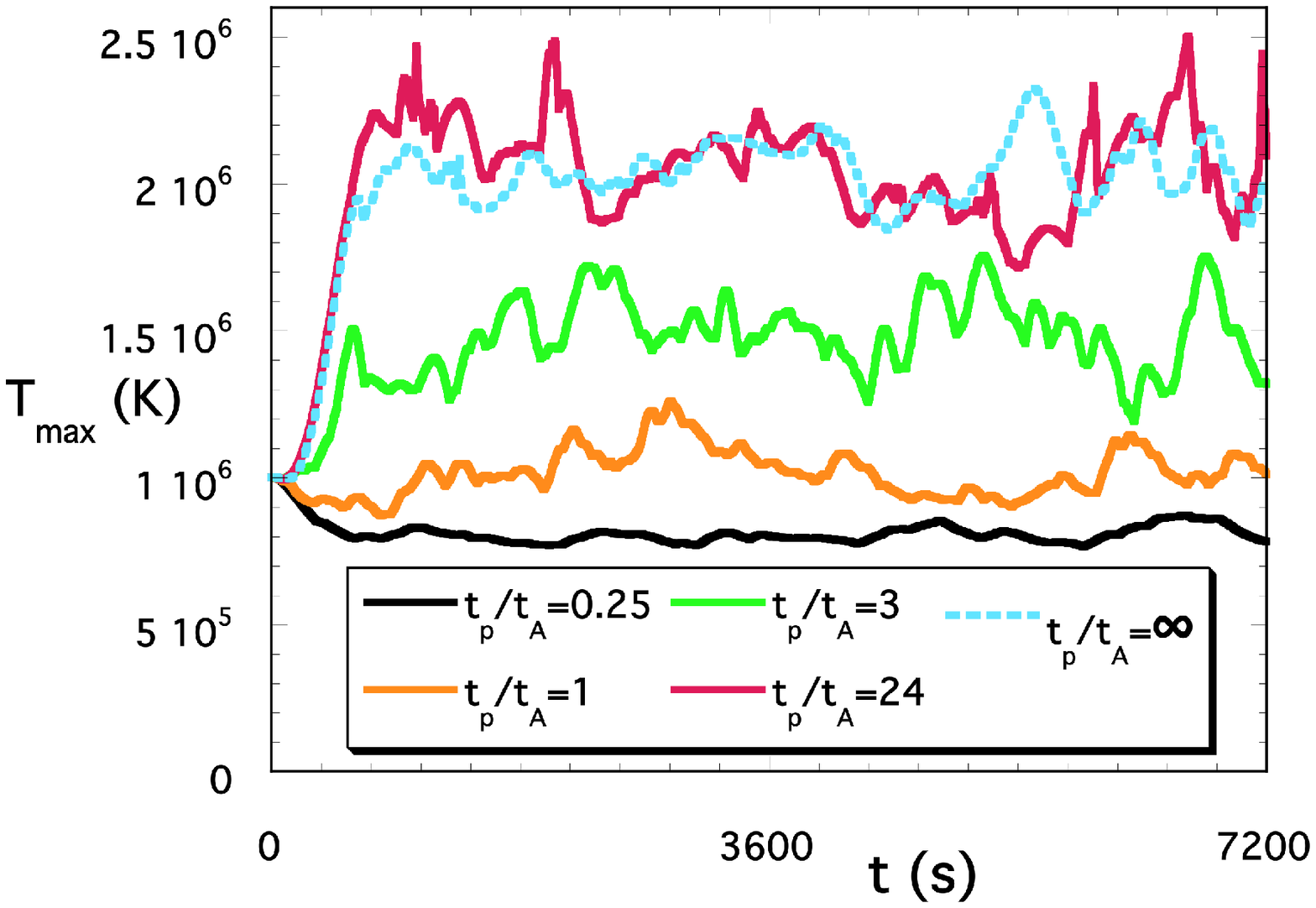} \hspace{1em}
		\includegraphics[width=.48\textwidth]{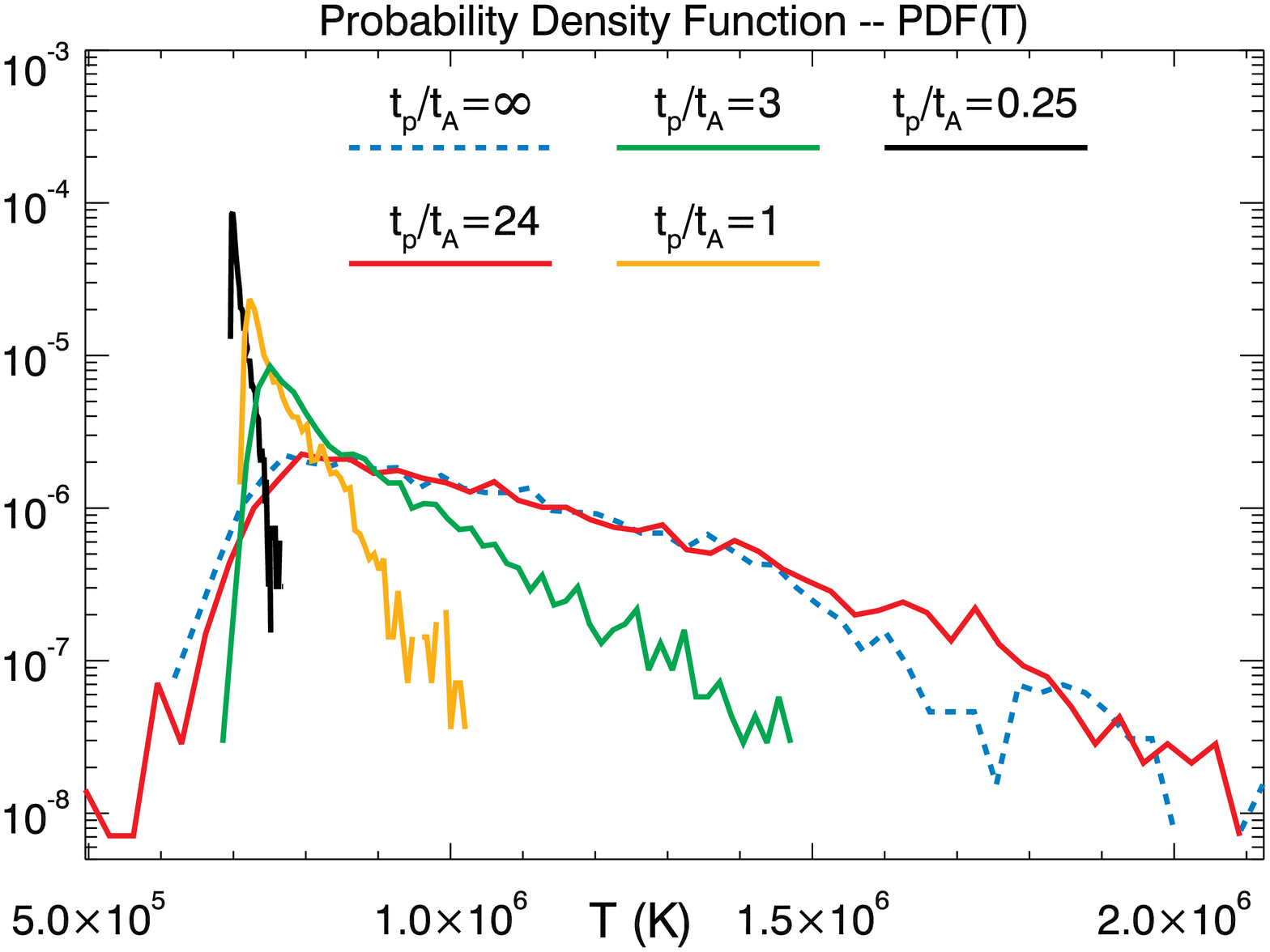}
	\end{center}
	\caption{\emph{Left panel:} Maximum temperature in the mid-plane, corresponding to the loop apex, as a function of time for different values of $t_p/t_A$. \emph{Right panel:} probability density function (PDF) of temperature T in the midplane at select time, showing the temperature plasma distribution at select times for same values of $t_p/t_A$ as in the left panel.}
	\label{fig:russ}
\end{figure*}

As mentioned earlier, in the simulations presented here we have changed the velocity variation time $t_p$ while keeping fixed the ratio $\ell_p/u_p=\tau_p$, the convective timescale, and the Alfv\'en crossing time $t_A=L/v_A$. In this way we have verified that higher frequencies yield significantly smaller heating rates for $t_p/t_A \lesssim$~1--3.  As discussed in Section~\ref{sec:da} in general we assume that the velocity variation timescale should be of the same order of the convective timescale $t_p \sim \tau_p$, and that in this case the non-dimensional field $\mathbf{u}$ and $\mathbf{b}$ solutions depending only on the single ratio $t_p/t_A$. Simulations carried out keeping $t_p$ fixed and changing $t_A$ are numerically more demanding, but \cite{2008ApJ...677.1348R} already discussed simulations with $\tau_p/t_A =$~1.25, 5, 25, 50, and 200 in the zero-frequency approximation. We found that also in this case the heating rates decreases strongly for smaller values of $\tau_p/t_A$, and that only the highest value of $\tau_p/t_A = 200$ has a high enough Poynting flux matching estimates for solar active regions. Additionally, as shown here, the zero-frequency case yields the highest heating rates, because fields decorrelation at the boundary is higher for higher frequencies. We can then conclude that smaller values of $t_p/t_A$ bring about smaller heating rates whether $t_p$ is changed independently of $\tau_p$ or not.

Compressible simulations discussed in the next section confirm the trends outlined by the reduced MHD simulations. In particular, as displayed in Figure~\ref{fig:russ}, the temperature profiles show a significantly decrease in maximum and average temperatures, with a consequently diminished emission in X-rays and EUV.

\section{Radiative Compressible MHD Simulations} \label{sec:cmhd}

To obtain information about the emissivity, it is necessary to perform simulations with a compressible MHD code. These independent simulations can also be used to confirm the trends outlined by the reduced MHD simulations. We use the HYPERION code to perform these simulations. HYPERION is a massively parallel code which solves the compressible MHD equations, including nonlinear thermal conduction and optically thin radiation \cite{2012A&A...544L..20D, 2016ApJ...817...47D}. The spatial model is the same as the one simulated by the RMHD code. Space is discretized with a hybrid Fourier collocation--finite difference scheme. Time step splitting is employed. A five-step fourth-order Runge-Kutta method is used for all terms except the thermal conduction, which is advanced using super time-stepping. Further detail can be found in \cite{2012A&A...544L..20D, 2016ApJ...817...47D}. Note that compressible MHD is more computationally intensive than RMHD. This limits both the computational times and Lundquist numbers for compressible MHD. Also, the CFL condition requires strict time step limitations due to the wide range of Alfv\'en speeds found in a coronal loop.

To employ compressible simulations requires us to be more specific about which star we are looking at. We need to choose a radiation function,  a value for gravitation, and relevant loop lengths and magnetic field strengths. HYPERION has been used extensively for solar coronal heating calculations. Hence the following calculations apply specifically to the solar case, but can be used as an example of a stellar case for which highly detailed observation are available. The simulations described here are for a coronal loop which is $25,000$\,km long, with a magnetic field strength of 0.01 Teslas. An initial temperature and number density are prescribed in which the temperature is set to $10^4$\,K at the upper chromospheric boundaries and $10^6$\,K at the loop apex \citep{2016ApJ...817...47D}.  The number density is set to achieve pressure balance. The simulations are run in dimensionless form. The relevant dimensionless parameters are: $S_v = \rho_0 V_A L_0/\mu \equiv$ viscous Lundquist number $=68958$, $S = \mu_0 V_A L_0/\eta \equiv$ Lundquist number $=88980$,  $\beta = \mu_0 p_0/B_0^2 \equiv$ pressure ration in the upper chromosphere $=1.7354\times 10^4$, $Pr = C_v \mu/\kappa T_0^{5/2} \equiv$ Prandtl number $=767$, and $P_{rad} = \mu/t_A^2 n_0^2 \log\Lambda(T_0) \equiv$ the radiative Prandtl number $=2.2222\times 10^{-14}$, that determines the strength of the radiation. $C_v$ is the specific heat at constant volume. The  The magnetic resistivity ($\eta$) and shear viscosity ($\mu$) are assumed to be constant and uniform, and Stokes relationship is assumed so the bulk viscosity $\lambda = 2/3 \mu$. The term ``$\kappa$'' denotes the thermal conductivity. Note that the Lundquist numbers and Prandtl numbers are scaled to permit the convergence of the numerical simulations.

In similar fashion to reduced MHD simulations $t_p/t_A$ is changed by varying $t_p$. This means that the simulations for various ratios all take about the same amount of time, whereas changing $t_A$ would make the high $t_A$ simulations much longer that the low ones. $t_A$ in dimensioned form is equal to $36.254$ seconds. Hence we find that the $t_p/t_A$ ratios of $0.01$, $0.25$, $1$, $3$, $24$ and $\infty$ correspond to $t_p = 0.36254$\,s,  $t_p = 9.0635$\,s, $t_p = 36.254$\,s, $t_p = 108.76$\,s, $t_p = 870.1$\,s, and $t_p = \infty$.

In the following, we consider the temporal and spatial consequences of changing the $t_p/t_A$ ratio on the thermodynamic variable in a solar coronal loop.

In particular, as displayed in Figure 4 (left panel), the temperature profiles show a significantly decrease in maximum and average temperatures, with a consequently diminished emission in X-rays and EUV.

The coronal loop, a magnetic configuration filled with plasma, is the physical structure that radiates X-rays. To gain further understanding we compute
the probability density function for the temperature  in the midplane of the computational box, corresponding to the loop apex, at select times, parametrized by $t_p/t_A$. As shown in the right panel of Figure~\ref{fig:russ} the temperature distributions is broad for $t_p/t_A \gtrsim 24$ with the maximum well beyond $10^6$\,K, while for smaller values of $t_p/t_A$ it becomes increasingly narrower with the bulk of the plasma at increasingly smaller temperatures. Since X-rays and EUV require temperatures higher than $10^6$\,K, the high-energy radiation emission decreases strongly for lower values of $t_p/t_A$ until vanishing for $t_p/t_A \sim$~ 1--3.

\section{Conclusions and Discussion}

In this paper we have investigated the dynamics of coronal loops forced by photospheric boundary motions to explore the heating of the magnetically confined regions of solar and stellar coronae.

A key parameter for the heating of coronal loops is the ratio of photospheric motions timescale $t_p$ to the axial Alfv\'en crossing time $t_A$. In fact in loops with $t_p/t_A \gtrsim$~10--24 photospheric motions inject Alfv\'en waves with wavelength much longer than the loop length. Consequently the dynamics and fields have a strong two-dimensional component, and high heating rates with plasma temperatures larger than a million K and a strong X-ray and EUV emission. On the other hand loops with $t_p/t_A \lesssim 3$ display a diminishing energy dissipation and temperature, and consequently fainter X-ray and EUV emission, the smaller the ratio $t_p/t_A$.

For all values of $t_p/t_A$  a statistically steady state is reached in which all physical quantities of interest fluctuate around a mean value. As in the case for long-timescale forcing, the dynamics for short-timescale forcings at $t_p/t_A \lesssim 1$ are characterized by a turbulent energy cascade that transfers energy from the large scales where it is injected by photospheric motions  to the small scales. The energy entering from the photospheric boundary per unit time $\epsilon_{in}=S$ (the integrated Poynting flux) is a key factor in determining the heating rate, together with the energy transfer rate across the inertial range $\epsilon_t$ and dissipation rate  $\epsilon_d$ (that in a steady state are all equal).
The reason for the decreased heating rates for smaller values of $t_p/t_A$ is the diminished Poynting flux.  The Poynting flux across the boundary $S_z = -B_0\ \mathbf{u_p} \cdot \mathbf{b}$ (Equation~(\ref{eq:poy})) depends indeed on the scalar product of the photospheric velocity at the boundary $\mathbf{u_p}$ and the magnetic field $\mathbf{b}$. As shown in Section~\ref{sec:la} during the linear stage the magnetic field is proportional to the boundary velocity field $\mathbf{b} \propto \mathbf{u}_p$ (Equations~(\ref{eq:linres1})-(\ref{eq:lincst1})). But for the case with $t_p/t_A = \infty$, where the fields have a strong two-dimensional component, as soon as nonlinearity sets in and the dynamics become turbulent the correlation between the magnetic field and the photospheric forcing decreases from 1 to about 0.65 for a boundary shear forcing velocity \citep{2010ApJ...722...65R} and from 1 to about 0.5 for photospheric granule-mimicking vortices (Figure~\ref{fig:corr}, see also Section~\ref{sec:bic}), because the turbulent dynamics alter the structure of the magnetic field decreasing the correlation. In the case of smaller values of $t_p/t_A$ the magnetic field has its energy in higher parallel modes (the smaller $t_p/t_A$). In this case the correlation between magnetic and velocity fields at the boundary has to remain high for a non vanishing Poynting flux, and as shown in Section~\ref{sec:la} this can only happen at resonant frequencies and depends critically on the sinusoidal shape of the wave that has to remain in phase with the boundary velocity forcing. But evidently the ensuing turbulent dynamics tends to destroy the wave-like structure of the magnetic field, possibly transferring energy in the m=0 mode, i.e., the two-dimensional component. In this way the magnetic field gets increasingly decorrelated (Figure~\ref{fig:corr}) from the velocity field at the boundary that has its specific frequency independent of the turbulent dynamics developing in the coronal layer, thus strongly decreasing the energy flux entering into the computational box (Figure~\ref{fig:en}).

In the heating of solar and stellar coronae many physical parameters enter into play. Nevertheless very little attention has been dedicated to the effect of photospheric motions, and in particular of the ratio of the photospheric timescale over the Alfv\'enic crossing time $t_p/t_A$.
This effect can determine what fraction of the stellar corona has a strong X-ray and EUV radiative emission, since given the magnetic field intensity and topology there will be a specific range of loop lengths with strong emission, while the emission will quickly fade away outside this range. Therefore the fraction of the coronal volume with high radiative emission and the total radiative emission of the star will be strongly influenced by the range of values of $t_p/t_A$ for that star.

This phenomenology could also have a strong impact in understanding the X-ray emission in so-called Ultra Cool Dwarfs. Recent observations show a large spread for the ratio of X-ray over bolometric luminosity for stars with similar parameters, leading to the hypothesis that two different magnetic field topologies (and therefore two different dynamos) could be present for this class of stars \citep{2014ApJ...785...10C}. Our results support this conjecture, since clearly stars with different topologies would have different values of $t_p/t_A$ for their loops, and consequently two different radiative emission intensities. In particular the average volumetric heating for loops with $t_p/t_A \gtrsim$\,10--24 is given by $S_z/L \sim \ell_p u_p B_0^2/4\pi L^2$~erg\,cm$^{-3}$\,s$^{-1}$ (Eq.~(\ref{eq:vhstar})), where $B_0$ is the intensity of the axial magnetic field threading the loop (in Gaussian units), 
$L$ is the loop length, $\ell_p$ the photospheric granule length-scale, and 
$u_p$ the convective velocity. Therefore in general longer loops with same axial magnetic field strength would have a smaller average heating rate and thus smaller maximum temperature and weaker X-ray and EUV emission. Clearly for a more quantitative analysis more data about the properties of convective motions and magnetic topology across the main sequence and in particular UCD would be required, and more numerical simulations addressing this specific issue should be performed.

Additionally the increased neutrality of the photospheres of late M and L dwarfs reduces the heating rates, because the field lines can partially slip through the photospheric plasma (resulting probably in a smaller effective photospheric shuffling velocity) and therefore the energy flux entering the corona in progressively decreased the higher the neutral fraction
\citep{2002ApJ...571..469M, 2010ApJ...709..332B}. We plan to investigate also this effect in subsequent work.

Finally we would like to stress that here we have investigated the dynamics of Parker model for coronal heating in solar and stellar coronae neglecting the effects of high frequency waves in loops (so-called AC heating). Our investigation show that the heating of coronal loops due to the shuffling by photospheric motions of magnetic field line footpoints decreases as the timescale ratio $t_p/t_A$ decreases while still in a relatively low frequency regime, i.e., considering motions with $t_p/t_A \sim \lambda_{\parallel}/L \gtrsim$~1/4. The effects of higher frequency photospheric motions or of higher frequency waves injected in the corona by phenomena occurring in the chromosphere or elsewhere and propagating in inhomogenous loops is beyond the scope of this investigation.

\section*{Acknowledgements}

This research was supported in part by the National Aeronautics and Space Administration Grant No.~NNH17AE96I issued through the Heliophysics Grand Challenge Research Program, NRL contract No.~N00173-17-P-3044, and by the NASA Parker Solar Probe Observatory Scientist grant No.~NNX15AF34G. RBD was also supported by the Naval Research Laboratory 6.1 program.
Computing resources supporting this work were provided by the NASA High-End Computing (HEC) Program through the NASA Advanced Supercomputing (NAS) Division at the Ames Research Center and by NRL LCP\&FD. 




\bibliographystyle{mnras}








\bsp	
\label{lastpage}
\end{document}